\documentclass[sigconf]{acmart}

\AtBeginDocument{%
  \providecommand\BibTeX{{%
    \normalfont B\kern-0.5em{\scshape i\kern-0.25em b}\kern-0.8em\TeX}}}

\setcopyright{acmcopyright}
\copyrightyear{2022}
\acmYear{2022}
\acmDOI{10.1145/1122445.1122456}

\acmConference[ICPC 2022]{The 30th International Conference on Program Comprehension}{May 21--22, 2022}{Pittsburgh, PA, USA}
\acmPrice{15.00}
\acmISBN{978-1-4503-XXXX-X/18/06}

\usepackage{listing}
\usepackage{listings}
\usepackage{algorithmic}
\usepackage{array}
\usepackage[lined, algonl, ruled, boxed,vlined,linesnumbered]{algorithm2e} 
\usepackage{multirow}
\usepackage{makecell}
\usepackage{xspace}
\usepackage{tcolorbox}
\usepackage{pifont}
\usepackage{slashbox}
\usepackage{booktabs}
\usepackage{url}
\usepackage{amsmath}
\usepackage{subfig}
\usepackage{textcomp}
\usepackage{fancyvrb}

\newcommand{\tool}{\textsc{Seader}\xspace}

\newcommand{\exampleCount}{28\xspace}
\newcommand{\vulCount}{86\xspace}
\newcommand{\pCount}{10\xspace}
\newcommand{\pre}{95\%\xspace}
\newcommand{\rec}{72\%\xspace}
\newcommand{\fscore}{82\%\xspace}
\newcommand{\sampleCount}{71\xspace}
\newcommand{\repairCount}{77\xspace}

\newcommand{\templateCount}{21\xspace}

\newcommand{\codefont}[1]{\footnotesize{\texttt{#1}}\normalsize}
\newcommand{\insecure}{\textbf{\emph{I}}\xspace}
\newcommand{\secure}{\textbf{\emph{S}}\xspace}
\newcommand{\p}{\textbf{\emph{P}}\xspace}

\newcolumntype{L}[1]{>{\raggedright\let\newline\\\arraybackslash\hspace{0pt}}m{#1}}
\newcolumntype{C}[1]{>{\centering\let\newline\\\arraybackslash\hspace{0pt}}m{#1}}
\newcolumntype{R}[1]{>{\raggedleft\let\newline\\\arraybackslash\hspace{0pt}}m{#1}}

\newcommand{\cmark}{\ding{51}}%
\copyrightyear{2022}
\acmYear{2022}
\setcopyright{rightsretained}
\acmConference[ICPC '22]{30th International Conference on Program Comprehension}{May 16--17, 2022}{Virtual Event, USA}
\acmBooktitle{30th International Conference on Program Comprehension (ICPC '22), May 16--17, 2022, Virtual Event, USA}
\acmDOI{10.1145/3524610.3527749}
\acmISBN{978-1-4503-9298-3/22/05}

\begin{document}

\title{Example-Based Vulnerability Detection and Repair in Java Code}

\author{Ying Zhang}
\email{yingzhang@vt.edu}
\affiliation{%
  \institution{Virginia Tech}
  \city{Blacksburg}
  \state{Virginia}
  \country{USA}
}

\author{Ya Xiao}
\email{yax99@vt.edu}
\affiliation{%
  \institution{Virginia Tech}
  \city{Blacksburg}
  \state{Virginia}
  \country{USA}
}

\author{Md Mahir Asef Kabir}
\email{mdmahirasefk@vt.edu}
\affiliation{%
  \institution{Virginia Tech}
  \city{Blacksburg}
  \state{Virginia}
  \country{USA}
}

\author{Danfeng (Daphne) Yao}
\email{danfeng@vt.edu}
\affiliation{%
  \institution{Virginia Tech}
  \city{Blacksburg}
  \state{Virginia}
  \country{USA}
}

\author{Na Meng}
\email{nm8247@vt.edu}
\affiliation{%
  \institution{Virginia Tech}
  \city{Blacksburg}
  \state{Virginia}
  \country{USA}
}

\begin{abstract}

The Java libraries JCA and JSSE offer cryptographic APIs to facilitate secure coding. 
When developers misuse some of the APIs, their code becomes
 vulnerable to cyber-attacks. 
 To eliminate such vulnerabilities, people built tools to detect security-API misuses via pattern matching. However, most tools do not (1)  fix misuses or (2) allow users to extend tools' pattern sets.  
To overcome both limitations, we created \tool---an example-based approach to detect and repair 
security-API misuses. 
Given an exemplar ${\langle}$insecure, secure${\rangle}$ code pair, 
 \tool compares the snippets to infer any API-misuse template and corresponding fixing edit. Based on the inferred info, given a program, \tool performs inter-procedural static analysis to search for security-API misuses and to propose customized fixes. 

For evaluation, we applied \tool to \exampleCount ${\langle}$insecure, secure${\rangle}$ code pairs; \tool successfully inferred \templateCount unique API-misuse templates and related fixes. With these $\langle$vulnerability, fix$\rangle$ patterns, we applied \tool to a program benchmark that 
has \vulCount known vulnerabilities. 
 \tool detected vulnerabilities with \pre precision, \rec recall, and \fscore F-score. 
We also applied \tool to 100 open-source projects and manually checked \repairCount suggested repairs;  
76 of the repairs were correct. 
\tool can help developers correctly use security APIs.    
   
\end{abstract}

\ccsdesc[500]{Software and its engineering~Software maintenance tools}
\ccsdesc[500]{Security and privacy~Software security engineering}

\keywords{Vulnerability repair, pattern inference, inter-procedural analysis}

\maketitle

\section{Introduction}
{JCA (Java Cryptography Architecture~\cite{jca}) and JSSE (Java Secure Socket Extension~\cite{jsse})} are two cryptographic frameworks,  provided by the standard Java platform. 
These frameworks offer security APIs to ease developers' secure software development. For instance, some of the APIs support key generation and secure communication.  
However, these libraries are not easy to use for two reasons. First, some APIs have overly complicated usage that is poorly documented~\cite{green2016developers,Nadi:2016}. Second, developers lack the necessary cybersecurity training to correctly implement security features~\cite{developer-lack-skills,meng2018secure,too-few-cybersecurity}.
Prior work shows that developers misused security APIs~\cite{Fischer2017,Rahaman2019}, and thus introduced vulnerabilities into software ~\cite{fahl2012eve,georgiev2012most}. For instance, Fischer et al.~found that the security-API misuses posted on StackOverflow~\cite{so} were copied and pasted into 196,403 Android applications available on Google Play~\cite{Fischer2017}. 
Fahl et al.~\cite{fahl2012eve} and Georgiev et al.~\cite{georgiev2012most}  showed that such API misuses in software could be exploited by hackers to steal data (e.g., user credentials). 

\begin{table*}
\footnotesize
\centering
\caption{Comparison of \tool against the existing detectors for security-API misuses}\label{tab:compare}\vspace{-2.em}
\begin{tabular}{l| ccc| ccc| cc}
\toprule
\multirow{2}{*}{\textbf{Tool}} & \multicolumn{3}{c|}{\textbf{API-Misuse Representation}}& \multicolumn{3}{c|}{\textbf{Misuse-Matching Strategy}} & \multicolumn{2}{c}{\textbf{Output}} \\ \cline{2-9}
& Built-in Rule & Template &Other &Intra-procedural Analysis & Inter-procedural Analysis &Other & Misuse & Repair
\\ \toprule
CryptoLint~\cite{egele2013empirical} &\cmark & &  
   &  &\cmark &
   & \cmark \\ \hline
CDRep~\cite{ma2016cdrep} &\cmark & &
   & &\cmark & 
   &\cmark &\cmark\\ \hline
CogniCrypt~\cite{kruger2017cognicrypt} & &\cmark & 
  & &\cmark &
  &\cmark\\ \hline
CryptoGuard~\cite{Rahaman2019} &\cmark & & 
  & &\cmark & 
  & \cmark\\ \hline  
FindSecBugs~\cite{findsecbugs} &\cmark & & 
 & &\cmark & 
 &\cmark \\ \hline
Fischer et al.'s tool~\cite{Fischer2017}  & & &\cmark 
  &\cmark & &\cmark 
  &\cmark \\ \hline
SonarQube~\cite{sonarqube} &\cmark & & 
  & &\cmark & 
  &\cmark \\ \hline
VuRLE~\cite{ma2017vurle} & & \cmark &
  &\cmark & & 
  & \cmark &\cmark \\ \hline
 SecureSync~\cite{Pham2010}& & &\cmark   
  &\cmark & & 
  &\cmark
  \\ \bottomrule 
\textbf{\tool} & &\cmark &
  & & \cmark & 
  &\cmark &\cmark\\
 \bottomrule
\end{tabular}
\vspace{-1.5em}
\end{table*}

Existing tools are insufficient to help developers eliminate security-API misuses. Table~\ref{tab:compare} summarizes both capability and extensibility of the mainstream techniques, and compares the tools with our new approach \tool. As shown in the table, existing tools usually represent cryptographic API misuses as built-in  rules~\cite{egele2013empirical,ma2016cdrep,Rahaman2019,findsecbugs,sonarqube}; 
users cannot easily extend these tools to detect more API-related vulnerabilities.  As more security libraries emerge and evolve, we believe that vulnerability detectors should have good extensibility to keep their pattern sets of API-misuses up-to-date. 
Although CogniCrypt~\cite{kruger2017cognicrypt} offers a domain-specific language (DSL), CrySL~\cite{kruger2018crysl}, for users to prescribe the usage templates of cryptographic APIs, users need to spend lots of time learning CrySL and crafting templates. 
VuRLE~\cite{ma2017vurle} infers templates from user-provided code examples. However, its algorithm does not observe the unique characteristics of security API-misuses (e.g., using an integer within certain range); thus, VuRLE cannot always detect or fix misuses effectively. 
 
 Additionally, most existing tools 
 merely report misuses, without suggesting any customized fixes. When developers lack the cybersecurity knowledge to understand the reported misuses, they may continue making mistakes when trying to fix those issues independently~\cite{Tupsamudre20}. Although CDRep~\cite{ma2016cdrep} and VuRLE~\cite{ma2017vurle} can suggest customized fixes, they are separately limited by (1) the inextensible hardcoded pattern set and (2) the intra-procedural analysis adopted for template matching. Please refer to Section~\ref{sec:related} for more details.

\begin{figure}[h]
    \centering
    \vspace{-1.em}
    \includegraphics[width=\linewidth]{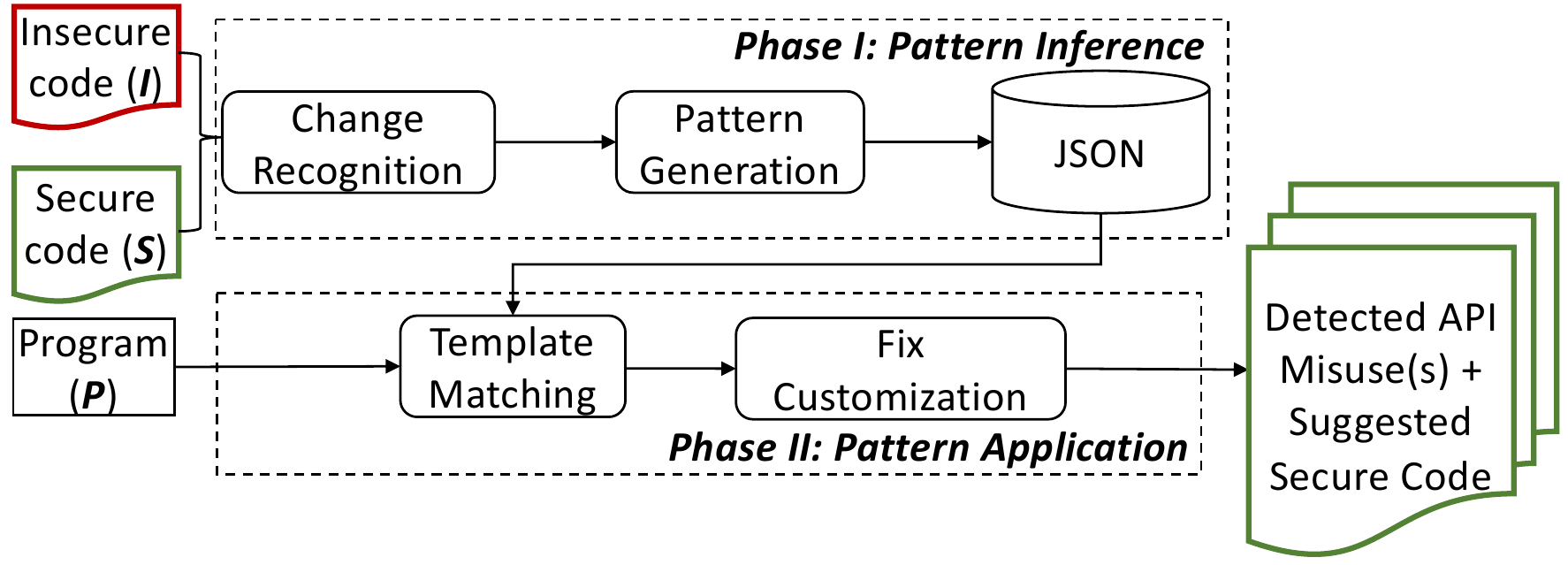}
    \vspace{-3.em}
    \caption{The overview of \tool}
    \label{fig:overview}
    \vspace{-1.em}
\end{figure}

To overcome the limitations of existing systems, we introduce \tool (short for ``\underline{se}curity-\underline{A}PI misuse \underline{de}tection and \underline{r}epair'')---our new approach for vulnerability detection and repair from a data-driven perspective.
 As shown in Figure~\ref{fig:overview}, there are two phases in \tool: pattern inference and pattern application. In Phase I, suppose that a domain expert (e.g., security researcher) provides

\vspace{-.1em}
\begin{itemize}
\item \emph{\textbf{I}}---insecure code with certain security-API misuse, and
\item \emph{\textbf{S}}---the secure counterpart showing the correct API usage.
\end{itemize}
\tool compares the two code snippets and detects program changes that can transform \emph{\textbf{I}} to \emph{\textbf{S}}. Next, based on those changes, \tool conducts \emph{intra-procedural} analysis to derive a vulnerability-repair pattern. Each pattern  
has two parts: (i) a vulnerable code template together with matching-related information, and (ii) the abstract fix. \tool stores all inferred patterns into a JSON file. 
In Phase II, given a program \textbf{\emph{P}}, \tool loads patterns from the JSON file, and conducts \emph{inter-procedural} program analysis to match code with any template. For each code match, 
\tool concretizes the corresponding abstract fix, and suggests code replacements  to developers. 

According to the existing API-misuse patterns mentioned in prior work~\cite{Fischer2017,Rahaman2019}, there are three unique kinds of security-API misuses that are hard to express with plain code examples, and are thus difficult to infer for existing program differencing-based approaches (e.g., VuRLE and SecureSync). Such misuses are about  API invocations with (i) constants instead of random values, (ii) multiple alternative specialized constants, or (iii) constants in certain value ranges (see Section~\ref{sec:edl} and Table~\ref{tab:classes} for more details). To facilitate users to describe these patterns via code examples, we defined three novel specialized ways of example specification, and developed \tool to specially infer patterns from those examples.  

For evaluation, we crafted \exampleCount $\langle$insecure, secure$\rangle$ code pairs based on the API-misuse patterns summarized by prior research. After \tool inferred patterns from those pairs, we further applied \tool to two program datasets to evaluate its effectiveness in  vulnerability detection and repair.  
When applied to the first dataset, \tool detected vulnerabilities with \pre precision, \rec recall, and \fscore F-score. 
After applying \tool to the second dataset, we inspected \repairCount repairs output by \tool and found 76 of them correct. 

To sum up, we made the following research contributions: 
\begin{itemize}
   \item We developed \tool---a new approach that performs \emph{intra-procedural analysis} to 
   infer vulnerability-repair patterns from $\langle$insecure, secure$\rangle$ code examples, does \emph{inter-procedural analysis} to match code with vulnerability templates, and customizes abstract fixes to suggest repairs. No prior work combines intra- with inter-procedural analysis in such a way.
   \item \tool supports specialized ways of example specification, which enable users to define examples for API misuses related to arbitrary constant parameters, constant parameters within certain ranges, and alternative constants. No prior work has such speciality to strengthen the expressiveness of example-based pattern specification.
\item We conducted a comprehensive evaluation with \tool. We observed that for vulnerability detection, \tool achieved a higher F-score than three state-of-the-art tools. For repair suggestion, \tool achieved 99\% (76/77) accuracy.   
    \end{itemize}
\tool's extensibility is realized by its capability of inferring patterns from provided $\langle I, S \rangle$ code examples. As security experts offer examples for new misuse patterns, \tool can infer those patterns to extend its pattern set. 
Additionally, \tool repairs misused APIs 
 by applying the inferred knowledge to given codebases.
We open-sourced our program and datasets at 
https://github.com/NiSE-Virginia-Tech/ying-ICPC-2022.

\vspace{-0.5em}
\section{A Motivating Example}
\label{sec:motivate}
This section overviews our approach with several code examples. 
Prior work shows that the security of symmetric encryption schemes depends on the secrecy of shared key~\cite{egele2013empirical}. Thus, developers should not generate secret keys from constant values hardcoded in programs~\cite{Fischer2017}. Suppose 
a security expert \textbf{Alex} wants to detect and fix such vulnerabilities using \tool. Alex needs to craft (1) an insecure code example to show the API misuse, and (2) a secure example for the correct API usage. As shown in Figure~\ref{fig:example}, the insecure code \insecure invokes the constructor of \codefont{SecretKeySpec} by passing in a constant array. Here, \codefont{ByteLiterals.CONSTANT\_ARRAY} is the specialized way that \tool requires users to adopt when they represent any byte-array constant.  Meanwhile, the secure code \secure invokes the same API with  \codefont{key}---a generated unpredictable value. 

\begin{figure}[h]
\footnotesize
\begin{tabular}{p{.98\linewidth}} \hline
\multicolumn{1}{c}{\textbf{Insecure code (\emph{I})}} \\ \hline
\vspace{-1em}
\begin{lstlisting}
void test() {
  SecretKey sekey= new SecretKeySpec(ByteLiterals.CONSTANT_ARRAY, "AES");  }
\end{lstlisting} 
\\ \hline
\multicolumn{1}{c}{\textbf{Secure code (\emph{S})}} \\ \hline
\vspace{-1em}
\begin{lstlisting}
// store the key as a field for reuse purpose
byte[] key = keyInit(); 

// create a key based on an unpredictable random value
public byte[] keyInit() {
  try {
    KeyGenerator keyGen=KeyGenerator.getInstance("AES");
    keyGen.init(256);
    SecretKey secretKey = keyGen.generateKey();
    byte[] keyBytes= secretKey.getEncoded();
    return keyBytes;
  } catch (Exception e) {
    e.printStackTrace();
    return null;
  }
}
void test() {
  SecretKey sekey= new SecretKeySpec(key, "AES"); }  
\end{lstlisting}\\ \hline
\end{tabular}
\vspace{-1.8em}
\caption{A pair of examples to show the vulnerability and repair relevant to secret key creation}\label{fig:example} 
\vspace{-1.9em}
\end{figure}

\begin{figure}[h]
\footnotesize
\begin{tabular}{p{.98\linewidth}} \hline
\multicolumn{1}{c}{\textbf{Vulnerable code template (T)}} \\ \hline
SecretKey \$v\_0\$ = new SecretKeySpec(ByteLiterals.CONSTANT\_ARRAY, "AES");
\\ \hline
\textbf{Matching-related data:}  \\
\hspace{1em}critical API: javax.crypto.spec.SecretKeySpec.SecretKeySpec(byte[], String) 

\hspace{1em}other security APIs: \{\}
\\ \hline
\multicolumn{1}{c}{\textbf{Abstract fix (F)}} \\ \hline
\textbf{Replace the matched statement with:}\\
\hspace{1em}SecretKey \$v\_0\$ = new SecretKeySpec(\$v\_1\$, "AES");\\
\textbf{Add these lines before the container method of the matched statement:}\\
\hspace{1em}// store the key as a field for reuse purpose\\
\hspace{1em}byte[] \$v\_1\$ = \$m\_0\$();\\
\hspace{1em}// create a key based on an unpredictable random value\\
\hspace{1em}public byte[] \$m\_0\$() \{\\
\hspace{2em}try \{\\
\hspace{3em}KeyGenerator \$v\_4\$=KeyGenerator.getInstance("AES");\\
\hspace{3em}\$v\_4\$.init(256);\\
\hspace{3em}SecretKey \$v\_3\$ = \$v\_4\$.generateKey();\\
\hspace{3em}byte[] \$v\_2\$= \$v\_3\$.getEncoded();\\
\hspace{3em}return \$v\_2\$;\\
\hspace{2em}\} catch (Exception \$v\_5\$) \{\\
\hspace{3em}\$v\_5\$.printStackTrace();\\
\hspace{3em}return null;\\
\hspace{2em}\}\\
\hspace{1em}\}\\ \hline
\end{tabular}
\vspace{-1.5em}
\caption{The pattern inferred from the code pair in Figure~\ref{fig:example}}\label{fig:pattern} 
\vspace{-2.em}
\end{figure}

Given the two examples, \tool generates abstract syntax trees (ASTs) and compares them for any AST edit operation.
For Figure~\ref{fig:example}, \tool creates an expression update and multiple statement insertions. The update operation replaces \codefont{ByteLiterals.CONSTANT\_}\codefont{ARRAY} with \codefont{key}. 
Next, based on the updated expression in \insecure, \tool conducts data-dependency analysis to find any security API that uses the expression, and treats it as a \textbf{critical API}. 
 Such critical APIs are important for \tool to later detect similar vulnerabilities in other codebases.
Afterwards, \tool generalizes
a \textbf{vulnerability-repair pattern} from the examples by abstracting away concrete variable/method names and edit-irrelevant code. 
As shown in Figure~\ref{fig:pattern}, the generalized pattern has two parts: 
 the vulnerability template (T) together with matching-related data, and  an abstract fix (F). 
Such pattern generalization ensures the transformation applicable
to codebases with distinct program contexts.

With a pattern inferred from the provided code pair, Alex can further apply \tool to an arbitrary program \p, to detect and fix any occurrence of the described vulnerability. In particular, given a program whose simplified version is shown in Listing~\ref{lst:p}, \tool first scans for any invocation of the critical API \codefont{SecretKeySpec(...)}. If no such invocation exists, \tool concludes that \p does not have the above-mentioned vulnerability; otherwise, if the API is invoked (see line 8 in Listing~\ref{lst:p}), \tool then searches for any code matching the template in Figure~\ref{fig:pattern}. The template-matching process  conducts inter-procedural analysis and checks for two conditions:

\begin{enumerate} 
\item[C1:] Is the first parameter derived from a constant? 
\item[C2:] Does the second parameter exactly match \codefont{"AES"}?
\end{enumerate}
If any invocation of \codefont{SecretKeySpec(...)} satisfies both conditions, \tool reports the code to be vulnerable. Notice that if we only check line 8 of Listing~\ref{lst:p}, neither \codefont{new String(passPhrase).} \codefont{getBytes()} nor \codefont{alg} satisfies any condition. Thanks to the usage of inter-procedural analysis, \tool can perform backward slicing to trace how both parameters are initialized. Because \codefont{alg} is a private field of \codefont{CEncryptor}, whose value is initialized on line 3 with \codefont{"AES"}, \tool decides that C2 is satisfied. Similarly, \codefont{passPhrase} is another field whose value is initialized with a parameter of the constructor \codefont{CEncryptor(...)} (lines 4-6). When \codefont{CEncryptor(...)} is called with parameter \codefont{"password"} before 
the invocation of \codefont{SecretKeySpec(...)} (lines 7-14), C1 is satisfied. Therefore, \tool concludes that line 8 matches the template; it matches concrete variable \codefont{secret} with  the template variable \codefont{\$v\_0\$}. 

\lstset{
numbers=left, 
basicstyle=\scriptsize,
breaklines=true,
language=java,
belowskip=8pt,
escapeinside={(*}{*)},
frame = tb
}

\begin{figure}
\vspace{-1em}
\begin{minipage}{\linewidth}
\begin{lstlisting}[caption=A simplified version of \p,label=lst:p]
public class CEncryptor {
  private char[] passPhrase;
  private String alg = "AES";
  public CEncryptor(String passPhrase) {
    this.passPhrase = passPhrase.toCharArray();
  }
  public Result encrypt(byte[] plain) throws Exception {
    SecretKey secret = new SecretKeySpec(new String(passPhrase).getBytes(), alg);
  ...
}
public class Main {
  public static void main(String[] args)
  CEncryptor aes0 = new CEncryptor("password");
  aes0.encrypt((byte[])args[0]);
  ...
}
\end{lstlisting}
\end{minipage}
\begin{minipage}{\linewidth}
\footnotesize
\begin{tabular}{p{.97\linewidth}}
\hline
\textbf{Replace the matched statement with:}\\
\hspace{1em}SecretKey secret = new SecretKeySpec(\$v\_1\$, "AES");\\
\textbf{Add these lines before the method encrypt(byte[] plain):}\\
1. \hspace{1em}// store the key as a field for reuse purpose\\
2. \hspace{1em}byte[] \$v\_1\$ = \$m\_0\$();\\
3. \hspace{1em}// create a key based on an unpredictable random value\\
4. \hspace{1em}public byte[] \$m\_0\$() \{\\
5. \hspace{2em}try \{\\
6. \hspace{3em}KeyGenerator \$v\_4\$=KeyGenerator.getInstance("AES");\\
7. \hspace{3em}\$v\_4\$.init(256);\\
8. \hspace{3em}SecretKey \$v\_3\$ = \$v\_4\$.generateKey();\\
9. \hspace{3em}byte[] \$v\_2\$= \$v\_3\$.getEncoded();\\
10.\hspace{3em}return \$v\_2\$;\\
11.\hspace{2em}\} catch (Exception \$v\_5\$) \{\\
12.\hspace{3em}\$v\_5\$.printStackTrace();\\
13.\hspace{3em}return null;\\
14.\hspace{2em}\}\\
15.\hspace{1em}\}\\ \hline
\end{tabular}
\vspace{-1.5em}
\caption{A customized fix for \p suggested by \tool}\label{fig:repair}
\vspace{-2.1em}
\end{minipage}
\end{figure}

\begin{figure*}
\centering
\includegraphics[width=0.87\textwidth]{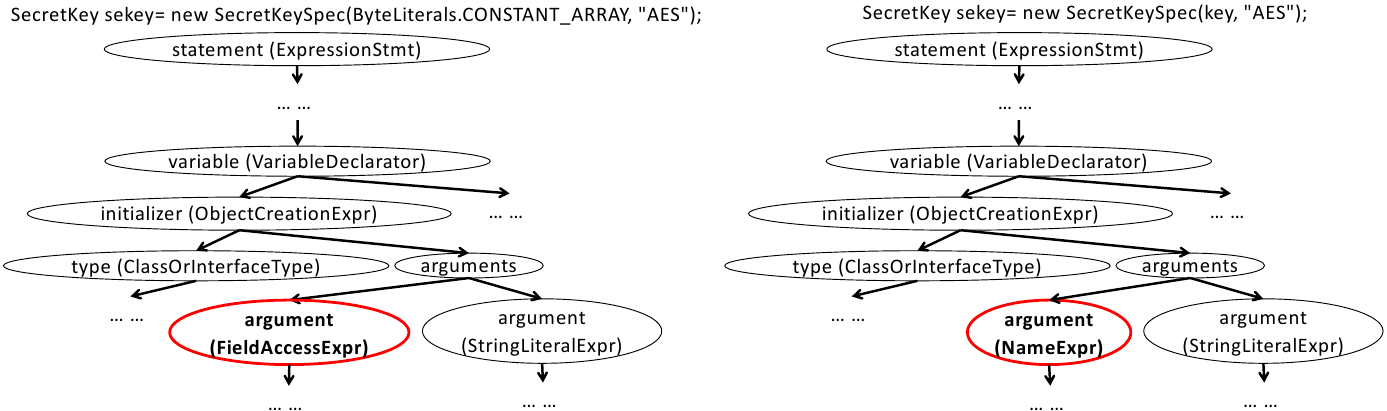}
\vspace{-1.em}
\caption{The simplified ASTs of the two statements related to a statement-level update operation}
\label{fig:asts}
\vspace{-1.5em}
\end{figure*}

For the found code match, \tool customizes the abstract fix shown in Figure~\ref{fig:pattern} by replacing the abstract variable \codefont{\$v\_0\$} with concrete variable \codefont{secret}. As shown in Figure~\ref{fig:repair}, the customized fix first initializes a \codefont{KeyGenerator} instance with the algorithm ``\codefont{AES}'' and the key size ``\codefont{256}'', to generate an unpredictable AES key (lines 6--8). Next, the AES key is converted to a byte array (line 9), which value can be stored into a Java field so that the value is reusable by both encryption and decryption modules. Additionally, inside the method \codefont{encrypt(...)}, the original vulnerable statement is updated to create a secret key using the generated byte array. 
\vspace{-0.5em}
\vspace{-0.5em}
\section{Approach}
\label{sec:approach}

There are two challenges to overcome in our research:
\begin{enumerate}
\item How can we infer generalized vulnerability-repair patterns from concrete $\langle$insecure, secure$\rangle$ code examples? 
\item How can we ensure that the inferred patterns are applicable to code 
  that is different from the original examples?
\end{enumerate} 
To address these challenges, 
as shown in Figure~\ref{fig:overview}, we designed two phases in \tool. The first phase takes two steps to infer vulnerability-repair patterns from $\langle$insecure, secure$\rangle$ code examples; the second phase contains another two steps to apply inferred patterns to given programs. In this section, we will first describe each of the four steps in detail (Section~\ref{sec:change}-Section~\ref{sec:fix}). 
Next, we will explain the three specialized ways of example specification, which can facilitate users to demonstrate certain API misuses (Section~\ref{sec:edl}). 

\vspace{-0.5em}
\subsection{Change Recognition}
\label{sec:change}
Given an $\langle I, S\rangle$ example pair, \tool compares code to locate (1) the root cause of any vulnerability demonstrated by $I$ and (2) the security patch shown in $S$. Specifically, \tool applies syntactic program differencing to the code pair, 
to reveal any edit operation(s) that can transform $I$ to $S$. This step consists of two parts: statement-level change recognition and expression-level change recognition. 

\vspace{-.3em}
\subsubsection{Statement-level change recognition} \tool first uses 
JavaParser~\cite{hosseini2013javaparser} to generate ASTs for $I$ and $S$, and then compares ASTs to create three types of edit operations: 

 \begin{itemize} 
\item \textbf{delete (Node $a$)}: Delete node $a$.

\item \textbf{insert (Node $a$, Node $b$, int $k$)}: Insert node $a$ and position it
as the $(k+1)^{th}$ child of node $b$.     

\item \textbf{update (Node $a$, Node $b$)}: Replace $a$ with
  $b$.  This operation changes $a$'s content. 
\end{itemize} 
Specifically, when comparing any two statements $s_i\in I$ and $s_s\in S$, \tool checks whether the code string of $s_i$ exactly matches that of $s_j$; if so, \tool considers $s_i$ unchanged while $I$ is transformed to $S$. Otherwise, if the code strings of $s_i$ and $s_j$ are different, \tool normalizes both statements by replacing concrete variables (e.g., \codefont{key}) with abstract ones (e.g., \codefont{\$v\_0}), and replacing constant values (e.g., \codefont{"AES"}) with abstract constants (e.g., \codefont{\$c\_0}). We denote the normalized representations as $n_i$ and $n_s$. Next, \tool computes the Levenshtein edit distance~\cite{levenshtein1966bcc} between $n_i$ and $n_s$, and computes the similarity score~\cite{fluri2007change} with: 

\vspace{-.5em}
\begin{equation*}
sim = 1 - \frac{edit\_distance}{max\_length(n_i, n_s)}
\end{equation*}  

The similarity score $sim$ is within [0, 1]. When $sim=1$, $n_i$ and $n_j$ are identical. We set a threshold $th=0.8$ such that if   $sim>=th$, $n_i$ and $n_j$ are considered to match. In this way, \tool can identify update operation(s). Compared with string-based match, the normalization-based match is more flexible, because it can match any  two statements that have similar syntactic structures but distinct variables or constants. 
Finally, if a statement $s_i\in I$ does not find a match in $S$, \tool infers a delete operation; if $s_s\in S$ is unmatched, \tool infers an insert operation.  

\vspace{-.3em}
\subsubsection{Expression-level change recognition}
For each statement-level update, \tool tries to identify any finer-granularity edit (i.e., expression replacement) to better comprehend changes, and to prepare for later pattern generation (see Section~\ref{sec:generalize}). When $s_i$ is updated to $s_s$, \tool conducts top-down matching between ASTs to identify edits. Namely, while traversing both trees in a preorder manner, \tool compares roots and inner nodes based on the AST node types, and compares leaf nodes based on the code content. Such node traversal and comparison continue until \tool finds all unmatched subtrees or leaves.

For the example code shown in Figure~\ref{fig:example}, with statement-level change recognition, \tool reveals one statement update and multiple statement additions. Figure~\ref{fig:asts} shows the simplified ASTs of both before- and after- versions for the updated statement. By comparing the ASTs in a top-down manner, \tool finds the first arguments sent to the constructor to differ (e.g., \codefont{FieldAccessExpr} vs.~\codefont{NameExpr}). Thus, \tool creates a finer-granularity operation to replace the statement-level update: \textbf{update (\codefont{ByteLiterals.CONSTANT\_ARRAY}, \codefont{key})}. 

Notice that we decided not to use existing tools, such as GumTree \cite{Falleri2014} and ChangeDistiller~\cite{fluri2007change},  to recognize changes for a variety of reasons. First, GumTree often mismatches nodes against developers' intent~\cite{Matsumoto2019}. GumTree can generate four types of edit operations: add, delete, update, and move. However, in our research, we need only three edit types: add, delete, and update, so that Seader can infer API-misuse patterns from recognized changes. Second, ChangeDistiller only detects statement-level changes, without identifying expression-level changes. Additionally, it also generates four edit types. 
 To avoid (1) fixing bugs in GumTree and (2) revising current tools to report three instead of four types of edit operations, we created our own program differencing algorithm.  

\vspace{-0.5em} 
\subsection{Pattern Generation}
\label{sec:generalize}
When security experts present an $\langle I, S\rangle$ example pair to demonstrate any API misuse, we expect that they provide the code snippets to show only one vulnerability and its repair. Additionally, based on our experience with security-API misuses, each vulnerability is usually caused by the misuse of one security API. Therefore, to infer a general vulnerability-repair pattern from a given code pair, we need to overcome two technical challenges:

\begin{itemize}
\item How can we identify the 
security API whose misuse is responsible for the vulnerability (i.e., critical API)? 
\item How should we capture any 
relationship between the critical API and its surrounding code?
\end{itemize}

\vspace{-.5em}
\subsubsection{Task 1: Identifying the critical API} Starting with the edit script $E$ created in Section~\ref{sec:change}, \tool looks for any update operation $update(e, e')$. If there is such an operation, \tool searches for the security API whose invocation is data-dependent on $e$ or $e'$, and considers the API to be \emph{critical}. For the example shown in Figure~\ref{fig:asts}, the critical API is \codefont{SecretKeySpec(byte[], String)} because it is invoked with the updated expression as the first argument.
Similarly, Figure~\ref{fig:example2} presents another example where a numeric literal is updated from \codefont{4} to \codefont{8}. With data-dependency analysis, \tool reveals that the constants are used to define variable \codefont{salt}, while \codefont{salt} is used as an argument when \codefont{PBEParameterSpec(...)} is invoked. Therefore, the method invocation depends on the updated expression, and the security API \codefont{PBEParameterSpec(byte[], int)} is considered \emph{critical}. 

\lstset{
numbers=left, 
breaklines=true,
language=java,
belowskip=-5pt,
escapeinside={(*}{*)},
frame=none
}

\begin{figure}
\footnotesize
\begin{tabular}{p{.97\linewidth}} \hline
\multicolumn{1}{c}{\textbf{Insecure code (\emph{I})}} \\ \hline
\vspace{-1em}
\begin{lstlisting}
void test(int iterations) {
  byte[] salt = new byte[4];
  AlgorithmParameterSpec paramSpec = new PBEParameterSpec(salt, iterations); }
\end{lstlisting} 
\\ \hline
\multicolumn{1}{c}{\textbf{Secure code (\emph{S})}} \\ \hline
\vspace{-1em}
\begin{lstlisting}
void test(int iterations) {
  byte[] salt = new byte[8];
  AlgorithmParameterSpec paramSpec = new PBEParameterSpec(salt, iterations); }
\end{lstlisting}\\ \hline
\end{tabular}
\vspace{-1.8em}
\caption{An $\langle I, S\rangle$ where a critical API \codefont{PBEParameterSpec(...)} indirectly depends on a  updated constant}\label{fig:example2} 
\vspace{-2.5em}
\end{figure}

If there is no update operation in $E$, \tool searches for any overridden security API that encloses all edit operations, and considers the overridden API to be \emph{critical}. Take the code pair shown in Figure~\ref{fig:example3} as an example. By comparing $I$ with $S$, \tool can identify one statement deletion and multiple statement insertions. As there is no update operation and all edit operations are enclosed by an overridden method \codefont{verify(String, SSLSession)} (indicated by \codefont{@Override}), \tool further locates the interface or super class declaring the method (e.g., \codefont{HostnameVerifier}). If the overridden method together with the interface/super class matches any known security API, \tool concludes the overridden method to be \emph{critical}. 

Lastly, if no update operation or overridden security API is identified, \tool checks whether there is any deletion of security API call in $E$; if so, the API is \emph{critical}. 
To facilitate later template matching (Section~\ref{sec:match}), for each identified critical API, \tool records the method binding information (e.g., \codefont{javax.crypto.spec.SecretKeySpec. SecretKeySpec(byte[], String)}). 

\begin{figure}
\footnotesize
\begin{tabular}{p{.97\linewidth}} \hline
\multicolumn{1}{c}{\textbf{Insecure code (\emph{I})}} \\ \hline
\vspace{-1em}
\begin{lstlisting}
public class HostVerifier implements HostnameVerifier {
  @Override
  public boolean verify(String hostname, SSLSession sslSession){
    return true;    }}
\end{lstlisting} 
\\ \hline
\multicolumn{1}{c}{\textbf{Secure code (\emph{S})}} \\ \hline
\vspace{-1em}
\begin{lstlisting}
public class HostVerifier implements HostnameVerifier {
  @Override
  public boolean verify(String hostname, SSLSession sslSession){
    //Please change "example.com" as needed
    if ("example.com".equals(hostname)) { 
      return true;
    }
    HostnameVerifier hv = HttpsURLConnection.getDefaultHostnameVerifier(); 
    return hv.verify(hostname, sslSession);    }}
\end{lstlisting}\\ \hline
\end{tabular}
\vspace{-1.8em}
\caption{A pair of examples from which \tool infers the critical API to be an overridden method}\label{fig:example3} 
\vspace{-2.5em}
\end{figure}

\vspace{-.3em}
\subsubsection{Task 2: Extracting relationship between the critical API and its surrounding code}
When a vulnerable code example has multiple statements (e.g., Figures~\ref{fig:example2} and~\ref{fig:example3}), we were curious how the critical API invocation is related to other statements. 
On one extreme, if the invocation is irrelevant to all surrounding statements, we should not include any surrounding code into the generalized pattern. On the other extreme, if the invocation is related to all surrounding code, we should take all code into account when inferring a vulnerability-repair pattern. Thus, this task intends to decide (1) which statements of $I$ to include into the vulnerable code template, (2) what additional security API call(s) to analyze for template matching (see Section~\ref{sec:match}), and (3) which statements of $S$ to include into the abstract fix. 

\tool performs intra-procedural data-dependency analysis. If a statement defines a variable whose value is (in)directly used by the critical API invocation, the statement is extracted as \emph{edit-relevant context}. \tool uses such context to characterize the demonstrated vulnerability. 
For the insecure code $I$ in Figure~\ref{fig:example2}, since the API call (line 3) data-depends on variable \codefont{salt}, lines 2-3 are extracted as context. Additionally, when the critical API is an overridden method, its code implementation in $I$ is considered edit-relevant context (see lines 3-4 in Figure~\ref{fig:example3}). Based on the extracted edit-relevant context, \tool abstracts all variables to derive a vulnerable code template $T$, and records mappings $M$ between abstract and concrete variables. 
In addition to the critical API, \tool also extracts binding information for any other security API invoked by the contextual code. Compared with edit-relevant context, these APIs provide more succinct hints. In our later template-matching process, these APIs can serve as ``\emph{anchors}'' for \tool to efficiently decide whether a program slice
is worth further comparison with the template. 

To locate the fix-relevant code in secure version $S$, \tool identifies any unchanged code in the edit-relevant context, the inserted statements, and the new version of any updated statement. For the secure code $S$ shown in Figure~\ref{fig:example2}, lines 2-3 are fix-relevant, because line 2 is the new version of an updated statement and line 3 is unchanged contextual code. Similarly, for the secure code $S$ shown in Figure~\ref{fig:example3}, lines 3-9 are fix-relevant, because lines 3 presents the critical API while lines 4-9 are inserted statements. Based on the above-mentioned variable mappings $M$ and fix-related code, \tool further abstracts variables used in the fix-related code to derive an abstract fix $F$. \tool ensures that the same concrete variables used in $I$ and $S$ are mapped to the same abstract variables. 

To sum up, given a $\langle I, S\rangle$ pair, \tool produces a pattern $Pat=\langle T, F\rangle$, which has a vulnerable code template $T$, an abstract fix $F$, and metadata to describe $T$ (i.e., bindings of security APIs). 

\vspace{-.5em}
\subsection{Template Matching}
\label{sec:match}

Given a program $P$, \tool uses a static analysis framework---WALA~\cite{wala}---to analyze the program JAR file (i.e., bytecode). 
As shown by lines 1.2--1.4 in Algorithm~\ref{alg1}, to find any code in $P$ that matches the template $T$, \tool first searches for the critical API (i.e., invocation or method reimplementation). If the critical API does not exist, \tool concludes that there is no match for $T$. Next, if the critical API is invoked at least once, for each invocation, \tool conducts inter-procedural backward slicing to retrieve all code $Sli$ on which the API call is data-dependent (i.e., \codefont{getBackwardSlice(x)}). When $T$ invokes one or more security APIs in addition to the critical API, \tool further examines whether $Sli$ contains matches for those extra APIs; if not, the matching trial fails (see lines 1.8--1.9). Next, \tool checks whether the matched code in $Sli$ preserves the data dependencies manifested by $T$ (i.e., \codefont{dataDependConsist(T, Sli)}).  
If those data dependencies also match, \tool reveals a vulnerability (see lines 1.10--1.11). 

Alternatively, if the critical API is reimplemented, for each reimplementation, \tool compares the code content against $T$, and reports a vulnerability if they match (see lines 1.13--1.14). At the end of this step, if any vulnerability is detected, \tool presents the line number where the critical API is invoked or is declared as an overridden method, and shows related matching details. The matching details include both code matches and abstract-concrete variable mappings. 

Actually, we designed our algorithm of template matching based on three considerations. First, as developers provide code examples in Java but WALA analyzes JAR files, template matching should leverage the minimum information (i.e., security APIs and variable data dependencies) to overcome any discrepancy between program representations (i.e., source code vs.~bytecode). Second, although \tool infers templates from simple code examples via intra-procedural analysis, we need to match code with templates step-by-step via inter-procedural analysis, so that \tool can find matches even if the program context is more complicated. Third, many security-API misuses are relevant to parameter usage or method overriding, so our matching algorithm observes such unique characteristics to establish matches.

\vspace{-1.5em}
\SetKwData{CandidateMatches}{Candi} 
\SetKwData{Matched}{Matched} 
\SetKwData{LeafMatches}{L} 
\SetKwData{SymbolicNameMapping}{S}

\begin{algorithm}
\caption{Matching Program P to template T} 
\label{alg1} 

\footnotesize
\KwIn{P, T, D \tcc{program, template, and related metadata}} 
\KwOut{Matched \tcc{a set of code matches from P to T}}

Candi $:=$ $\emptyset$, Matched $:= \emptyset$; \\
\tcc{1. search for matches of the critical API} 
\ForEach{code line x $\in$ P}{
  \If{x invokes D(critical) || x declares D(critical)}{ 
    Candi := Candi $\cup$ x; 	
  } 
}

\ForEach{x $\in $Candi}{
  \If{x invokes {\tt D}(critical)}{
\tcc{2(a). For API call, do program slicing and look for matches of other security APIs} 
Sli = getBackwardSlice(x);\\
\If{(Sli has all matches for D(other)) == false}{
continue;  
}
\tcc{3. check whether the data dependencies between security APIs in T match those in Sli}
\If{dataDependConsist(T, Sli)}{
Matched:=Matched $\cup$ \{Sli, mappings\};
}
}\Else{
\tcc{2(b). For API overriding, check the code}
\If{contentMatch(code(P, x), T)}{
Matched := Matched $\cup$ \{code(P, x), mappings\};
}
}
}
\vspace{-0.6em}
\end{algorithm} 

\vspace{-.5em}
\subsection{Fix Customization}
\label{sec:fix}
This step involves two types of customization: variable customization and edit customization. To customize variables, based on the matching details mentioned in Section~\ref{sec:match}, \tool replaces abstract variables in $F$ with the corresponding concrete ones. 
We denote this customized version as $F_c$. For edit customization, \tool suggests code replacements in two distinct ways depeding on the inferred edit operations mentioned in Section~\ref{sec:change}. Specifically, if there is only one update operation inferred, \tool simply recommends an alternative expression to replace the original expression. Otherwise, \tool presents $F_c$ for developers to consider.

Notice that \tool does not directly modify $P$ to repair any vulnerability for two reasons. First, when template $T$ contains multiple statements, it is possible that the corresponding code match involves statements from multiple method bodies. Automatically editing those statements can be risky and cause unpredictable impacts on program semantics. 
Second, some fixes require for developers' further customization  based on their software environments (e.g., network configurations, file systems, and security infrastructures). As implied by Figure~\ref{fig:example3}, the abstract fix derived from $S$ will contain a comment \codefont{"//Please change 'example.com' as needed"}, so will the customized fix by \tool. This comment instructs developers to replace the standard hostname based on their circustances. 

\vspace{-0.5em}
\subsection{Specialized Ways of Example Specification}
\label{sec:edl}
We believe that by crafting $\langle I, S\rangle$ code pairs, security experts can demonstrate the misuse and correct usage of security APIs. However, we also noticed some scenarios where plain Java examples cannot effectively reflect the vulnerability-repair patterns. To solve this problem, we defined three \textbf{stub Java classes} (i.e., fake classes) for user adoption and invented \textbf{three specialized ways of example definition}. 
As shown in Table~\ref{tab:stubs}, the stub classes offer stub methods or fields to facilitate constant-related example specification. This section explains the scenarios where our special specification methods are needed. 

\begin{table}
\footnotesize
\centering
\caption{The stubs defined to ease example specification}\label{tab:stubs}
\vspace{-1.8em}
\begin{tabular}{p{1.2cm}|p{1.5cm}|p{4.7cm}}
\toprule
\textbf{Class} & \textbf{Members} & \textbf{Semantics} \\ \toprule
& {StringLiterals (String... a)} &This constructor creates a {\tt StringLiterals} object with one or more string literals.\\ \cline{2-3}
{StringLiterals}  & getAString() & This method randomly returns one of the strings originally used to construct the {\tt StringLiterals} object.
\\ 
\hline
ByteLiterals & CONSTANT\_ ARRAY & This field serves as a placeholder for a byte-array constant, whose value can be unspecified.\\ \hline
CharLiterals & CONSTANT\_ ARRAY & This field serves as a placeholder for a char-array constant, whose value can be unspecified.\\ 
\bottomrule
\end{tabular}
\vspace{-2.5em}
\end{table}

\vspace{-.5em}
\paragraph{Scenario 1. An API misuse involves an \textbf{arbitrary} constant value instead of any \textbf{particular} constant} Plain examples only show the usage of particular constant values, but cannot generally represent the constant concept. Consider the vulnerability introduced in Section~\ref{sec:motivate}. Without using  \codefont{ByteLiterals.CONSTANT\_ARRAY}, a domain expert  
has to define a plain example to show the API misuse, such as 

\codefont{SecretKey sekey = new SecretKeySpec("ABCDE".getBytes(), "AES");}

\noindent
\tool is designed to preserve all string literals from $I$ when generalizing template $T$, and to look for those values when matching code with $T$. Consequently, given the above-mentioned example, \tool will inevitably embed \codefont{"ABCDE"} into the inferred template. 
To help users avoid such unwanted literal values in $T$,  we defined \codefont{ByteLiterals.CONSTANT\_ARRAY} and \codefont{CharLiterals.CONSTANT\_ARRAY}. These static fields can be used as placeholders or wildcards for constant arrays, to represent the general constant concept in examples. 
When \tool detects such fields in examples, it keeps them  as they are in $T$ and later matches them with constant values in $P$.

\vspace{-.3em}
\paragraph{Scenario 2. An API misuse has multiple alternative insecure (or secure) options} Given a parameter of certain security API, suppose that there are (1) $m$ distinct values to cause API misuse and (2) $n$ alternatives to ensure correct API usage, where $m\ge1$, $n\ge1$. To express all possible combinations between the insecure and secure options via plain Java examples, users have to provide $m\times n$ pairs of examples, which practice is inefficient and undesirable. To solve this issue, we defined two stub methods in \codefont{StringLiterals}. As shown in Figure~\ref{fig:multi}, one is a constructor of \codefont{StringLiterals}, which can take in any number of string literals as arguments (see line 1 in $I$) and store those values into an internal list structure. The other method is \codefont{getAString()}, which randomly picks and returns a value from that list (see line 2 in $I$). In this way, a domain expert can efficiently enumerate multiple secure/insecure options in just one code pair. 

\begin{figure}
\scriptsize
\begin{tabular}{p{.97\linewidth}} \hline
\multicolumn{1}{c}{\textbf{Insecure code (\emph{I})}} \\ \hline
\vspace{-1em}
\begin{lstlisting}
StringLiterals literals=new StringLiterals("AES", "RC2", "RC4", "RC5", "DES", "blowfish", "DESede", "ARCFOUR");
Cipher.getInstance(literals.getAString());
\end{lstlisting} 
\\ \hline
\multicolumn{1}{c}{\textbf{Secure code (\emph{S})}} \\ \hline
\vspace{-1em}
\begin{lstlisting}
StringLiterals literals = new StringLiterals("AES/GCM/NoPadding","RSA/ECB/OAEPWithSHA-1AndMGF1Padding");
Cipher.getInstance(literals.getAString());
\end{lstlisting}\\ \hline
\end{tabular}
\vspace{-2.em}
\caption{A code pair where multiple alternative secure and insecure options are specified simultaneously}\label{fig:multi} 
\vspace{-2.5em}
\end{figure}

The examples in Figure~\ref{fig:multi} show that when security API
 \codefont{Cipher. getInstance(...)} is called, the parameter may have one of the insecure values (e.g., \codefont{"AES"}). Such vulnerability can be addressed when the value is replaced by one of the three secure options (e.g., \codefont{"AES/GCM/NoPadding"}). 
 Given the example in Figure~\ref{fig:multi}, \tool extracts insecure and secure options from \codefont{StringLiterals}-related statements, detects vulnerabilities in $P$ if the security API is invoked with any insecure option, and suggests
all secure alternatives.   

\vspace{-.3em}
\paragraph{Scenario 3. An API misuse requires for a parameter value in a specific range} Given an integer parameter $p$ of certain API, suppose that there is a threshold value $th$ such that the API invocation is secure only when $p\ge th$. To enumerate all possible vulnerable cases and related repairs via plain examples, theoretically, a user has to provide $(th-Integer.MIN\_VALUE) \times (Integer.MAX\_VALUE-th+1)$ code pairs, which practice is infeasible. Therefore, we invented a special way of example definition, which requires users to provide only (1) one insecure example by setting $p$ to a concrete value less than $th$ and (2) one secure example by setting $p=th$. 
As shown in Figure~\ref{fig:example2}, if a security expert wants to describe the pattern that \emph{the array size of the first parameter should be no less than 8}, then s/he can define $I$ by creating an array with a smaller size (i.e., 4) and define $S$ by setting the size to 8. 
\tool can identify the integer literals used by $I$ and $S$, and infer the secure value range $size\ge 8$. 

\vspace{-0.5em}
\section{Evaluation}

This section first describes the evaluation datasets and metrics, and then presents \tool's effectiveness of pattern inference. Next, it explains the tool effectiveness of pattern application, including vulnerability detection and repair. 
We did all experiments on Linux Mint 20.3 Cinnamon, version 5.2.7; we used Intel Core i7-8700 processor and 32GB memory.

\vspace{-0.5em}
\subsection{Datasets}
\label{sec:datasets}

We used one dataset to evaluate pattern inference, and two datasets to evaluate pattern application.

\begin{table*}
\centering
\caption{The API misuses and related fixes summarized by prior work~\cite{Mendel2013,tls1.2,Fischer2017,Rahaman2019}}\label{tab:classes}
\vspace{-1.5em}
\footnotesize
\begin{tabular}{R{0.3cm}|p{2.cm}|p{7.2cm}|p{6.8cm}}
\toprule
\textbf{Id} & \textbf{Security Class API} & \textbf{Insecure} & \textbf{Secure}\\ \toprule
1 & Cipher & The algorithm and/or mode is set as AES, RC2, RC4, RC5, DES, DESede, AES/ECB, Blowfish, ARCFOUR, or RSA/None/NoPadding. & The algorithm and/or mode is set as AES/GCM/NoPadding, RSA/ECB/OAEPWithSHA-1AndMGF1Padding, AES/CFB/PKCS5Padding, or RSA/CBC/PKCS5Padding.  \\ \hline 
2 & {HostnameVerifier} & Allow all hostnames. & Implement logic to actually verify hostnames.  \\ \hline
3 &  {IvParameterSpec} & Create an initialization vector (IV) with a constant. & Create an IV with an unpredictable random value. \\ \hline
4 &  {KeyPairGenerator} & Create an RSA key pair where key size $<$ 2048 bits or create an ECC key pair where key size $<$ 224 bits. & RSA key size $>=$ 2048 bits, ECC key size $>=$ 224 bits. \\ \hline
5 &  {KeyStore} & When loading a keystore from a given input stream, the provided password is a hardcoded constant non-null value. & The password is retrieved from some external source (e.g., database or file).  \\ \hline
6 &  {MessageDigest} & The algorithm is MD2, MD5, SHA-1, or SHA-224. & The algorithms is SHA-256, SHA-512 or SHA-3.\\ \hline
7 &  {PBEKeySpec} & Create a PBEKey based on a constant salt. & Use an unpredictable random salt value to create the key.\\ \hline
8 &  {PBEParameterSpec} & Create a parameter for password-based encryption (PBE) by setting salt size $<$ 64 bits or iteration count $<$ 1000, or by using a constant salt. & Salt size $>=$ 64 bits, iteration count $>=$1000. Use an unpredictable randomly generated salt value. \\ \hline
9 &  {SecretKeyFactory} & Create secret keys with algorithm DES, DESede, ARCFOUR, PBEWithMD5AndDES, or PBKDF2WithHmacSHA1. & Create secret keys with AES or PBEWithHmacSHA256AndAES\_256.\\ \hline
10 &  {SecretKeySpec} & Create a secret key with a constant value, or using the algorithm DES, DESede, Blowfish, HmacSHA1, ARCFOUR, PBEWithMD5AndDES, or PBKDF2WithHmacSHA1. & Create a secret key with an unpredictable randomly generated value, or using the algorithm AES or PBEWithHmacSHA256AndAES\_128.  \\ \hline
11 &  {SecureRandom} & Use Random to generate random values, or set SecureRandom to use a constant seed. & Use SecureRandom instead of Random, and ensure the seed to be a random value. \\ \hline
12 &  {SSLContext} & Use the protocol SSL, TLSv1.0, or TLSv1.1. & Use the protocol TLSv1.2 or TLSv1.3\\ \hline
13 &  {TrustManager} & Trust all clients or servers & Check clients and/or check servers. \\ \bottomrule
\end{tabular}
\vspace{-2.em}
\end{table*}

\vspace{-.3em}
\subsubsection{A dataset to evaluate pattern inference}
\label{sec:evaluateInfer}

Prior research revealed a number of security-API misuses and related correct usage in Java~\cite{rfc7525,rfc2898,Manger2001,fahl2012eve,Mendel2013,tls1.2,Fischer2017,meng2018secure,chen2019reliable,Rahaman2019}. To evaluate \tool's effectiveness of pattern inference, we referred to those well-described API misuses and fixes while crafting code examples for \tool. Table~\ref{tab:classes} lists the 13 security class APIs we focused on, the insecure usage of certain method API(s) frequently mentioned by prior work, and the secure usage. With this domain knowledge, we handcrafted \exampleCount $\langle I, S\rangle$  pairs. Among the pairs, 19 pairs are defined in the specialized ways introduced in Section~\ref{sec:edl}, and 9 pairs are defined with plain Java examples. Within the 19 pairs, 8 pairs, 6 pairs, and 5 pairs separately belong to Scenarios 1--3.

\vspace{-.3em}
\subsubsection{Two datasets to evaluate pattern application}
\label{sec:evaluateApply} 

The first dataset is a third-party benchmark, consisting of \vulCount real vulnerabilities from \pCount Apache open-source projects~\cite{apache-bench,afrose2021evaluation}. 
We decided to use this dataset for two reasons. First, it was created by other researchers, so it can be used to objectively assess the effectiveness of different vulnerability detectors.
Second, 
most of the \vulCount vulnerabilities belong to the 13 security classes shown in  Table~\ref{tab:classes}, so 
they can properly measure \tool's capability of pattern application.  
The second dataset contains 100 widely used Apache open-source projects. To create this dataset, we first ranked the Apache projects available on GitHub~\cite{gh} in a descending order of their popularity (i.e., star counts). 
Next, we located the top 100 projects that satisfy the following constraints: (1) the project uses the security APIs that \tool examines; (2) the project is compilable because \tool analyzes the compiled JAR files. 
The resulting dataset is used to evaluate \tool's effectiveness of repair suggestion. 

\vspace{-0.5em}
\subsection{Metrics}
\label{sec:metrics}

As with prior work~\cite{Rahaman2019}, we leveraged the following three metrics to measure tools' capability of vulnerability detection: 

\emph{\textbf{Precision (P)}} measures among all reported vulnerabilities, how many of them are true vulnerabilities. 

\vspace{-0.5em}
\begin{equation*}
P=\dfrac{\mbox{\# of correct reports}}{\mbox{Total \# of reports}} 
\end{equation*}
\normalsize

\noindent
When a tool reports a set of vulnerabilities $S_1$ and the known set of vulnerabilities is $S_2$, we intersected $S_1$ with $S_2$ to automatically compute precision. Namely, $P=|S_1\cap S_2|/|S_1|$.

\emph{\textbf{Recall (R)}} measures among all known vulnerabilities, how many of them are detected by a tool. 

\vspace{-0.5em}
\begin{equation*}
R=\frac{\mbox{\# of correct reports}}{\mbox{Total \# of known vulnerabilities}}
\end{equation*}

\noindent
When a tool reports a set of vulnerabilities $S_1$, we intersected $S_1$ with the set of known vulnerabilities $S_2$ to automatically compute recall, i.e., $R=|S_1\cap S_2|/|S_2|$.

\emph{\textbf{F-score (F)}} is the harmonic mean of precision and recall; it can reflect the trade-off between precision and recall. 

\vspace{-0.5em}
\begin{equation*}
F = \frac{2 \times P \times R}{P + R} 
\end{equation*}
\vspace{-2em}
\vspace{-0.5em}
\subsection{Effectiveness of Pattern Inference}
\label{sec:infer}

As mentioned in Section~\ref{sec:datasets}, we crafted \exampleCount code pairs to evaluate \tool's effectiveness of pattern inference. We categorized the \exampleCount pairs based on two criteria: 
\begin{enumerate}
\item[C1.] Do $I$ and $S$ contain single or multiple statements? 
\item[C2.] Does pattern inference abstract variables? 
\end{enumerate}
The two conditions actually reflect the difficulty levels or challenges of these pattern inference tasks. For instance, if $I$ or $S$ has multiple statements, \tool conducts data-dependency analysis to locate the edit-relevant context in $I$ or to reveal the fix-relevant code in $S$. If $I$ or $S$ uses variables, \tool abstracts all variable names to ensure the general applicability of inferred patterns. As shown in Table~\ref{tab:examplesInfer}, there are four simplest pairs; \tool can handle these pairs without conducting any data-dependency analysis or identifier generalization. Meanwhile, there are 18 most complicated cases that require \tool to analyze data dependencies and generalize identifiers. 
 In our evaluation, \tool correctly inferred patterns from all pairs. When some pairs present secure/insecure options (e.g., distinct string literals) for the \emph{same} critical API, \tool merged the inferred patterns.  
 In this way, \tool derived \templateCount unique patterns.

\begin{table}
\caption{The \exampleCount code pairs for pattern inference}\label{tab:examplesInfer}
\vspace{-1.5em}
\footnotesize
\centering
\begin{tabular}{R{1cm}|R{2.cm}|R{2.3cm}} \toprule
& \textbf{Single statement} & \textbf{Multiple statements} \\ \hline
\textbf{Identical} & 4 & 5\\ \hline
\textbf{Abstract} & 1 & 18 \\ 
 \bottomrule
\end{tabular}\vspace{-2.em}
\end{table}

\vspace{.2em}
\noindent\begin{tabular}{|p{8.1cm}|}
	\hline
	\textbf{Finding 1:} \emph{Our experiment shows \tool's great capability of pattern inference. \tool shows impressive extensibility by  inferring patterns from various examples.
	}
\\
	\hline
\end{tabular}

\vspace{-.5em}
\subsection{Effectiveness of Vulnerability Detection}
\label{sec:detect}

\begin{table*}
\centering
\footnotesize
\caption{Evaluation results on the \vulCount-vulnerability dataset~\cite{apache-bench}}\label{tab:32}
\vspace{-1.8em}
\begin{tabular}{c|R{1.7cm}| R{0.5cm}R{0.5cm}R{0.5cm}| R{0.5cm}R{0.5cm}R{0.5cm} | R{0.5cm}R{0.5cm}R{0.5cm} | R{0.5cm}R{0.5cm}R{0.5cm}} \toprule
\multirow{2}{*}{\textbf{Apache Project}} & \textbf{\# of Labeled}  & \multicolumn{3}{c|}{\textbf{CogniCrypt}} & \multicolumn{3}{c|}{\textbf{CryptoGuard}} &\multicolumn{3}{c|}{\textbf{FindSecBugs}} &\multicolumn{3}{c}{\textbf{\tool}}\\ \cline{3-14}
& \textbf{Vulnerabilities} & \textbf{P(\%)} & \textbf{R(\%)} & \textbf{F(\%)} & \textbf{P(\%)} & \textbf{R(\%)} & \textbf{F(\%)} & \textbf{P(\%)} & \textbf{R(\%)} & \textbf{F(\%)} & \textbf{P(\%)} & \textbf{R(\%)} & \textbf{F(\%)} \\ \toprule
deltaspike.jar & 2  
	& 40 & 100 & 57  
	& 100 & 100 & 100
	& 100 & 100 & 100
	& 100 & 100 & 100\\ \hline
directory-server.jar (apacheds-kerberos-codec) & 19  
	& 51 & 95 & 67  
	& 100 & 26  & 42 
	& 100 & 58  & 73
	& 94 & 84 & 89 \\ \hline 
incubator-taverna-workbench.jar & 5 
	& 57 & 80 & 67  
	& 100 & 80 & 89
	&100 & 40 & 57
	& 80 & 80 & 80 \\ \hline
manifoldcf.jar (mcf-core) & 3  
	& 17 & 33 & 22 
	& 60 & 100 & 75
	& 75 & 100 & 86  
	& 75 & 100 & 86 \\ \hline
meecrowave.jar & 3  
	& 100 & 100 & 100 
	& 100 & 67 & 80
	& 100 & 67 & 80
	& 100 & 100 & 100 \\ \hline
spark.jar & 27  
	&100 &26 & 41  
	& 100 & 100 & 100 
	& 100 & 85 & 92
	& 100 & 93 & 96\\ \hline
tika.jar & 0 
	& - & - & - 
	& - & - & -
	& - & - & -
	& - & - & - \\ \hline
tomee.jar (openejb-core) & 7 
	& 60 & 43 & 50 
	& 83 & 71 & 77 
	&60 & 43 & 50
	& - & - & - \\ \hline
wicket.jar &5 
	& 40 & 40 & 40
	& 100 &100&100
	& 100 & 40 & 57
	&100 & 60 &75\\ \hline
artemis-commons.jar &15
	&100 &40 &57 
	&100 &27 &42
	&100 &33&50 
	&100 &40 &57\\ 
\bottomrule
\textbf{Overall} & 86 
	& 58 & 53  & 56 
	& 95 & 66  & 78 
	& 95 & 62 & 75
	& 95 & 72 & 82 \\ \bottomrule
\multicolumn{14}{l}{``-'' means no value is computed,  because there is no labeled API misuse in the ground truth dataset or there is no tool-reported misuse.}	
\end{tabular}
\vspace{-2.em}
\end{table*}

To assess \tool's capability of vulnerability detection, we used a third-party dataset (see Section~\ref{sec:evaluateApply}). We applied \tool and three state-of-the-art vulnerability detectors (i.e., CogniCrypt~\cite{kruger2017cognicrypt}, CryptoGuard~\cite{Rahaman2019}, and FindSecBugs~\cite{findsecbugs}) to all subject programs.
The tool versions we used include CogniCrypt-2.7.1, commit 94135c5 of CryptoGuard, and findsecbugs-cli-1.10.1. 
To ensure that CogniCrypt has enough memory during execution, instead of using its default configuration, we set the maximum heap size to 30G (-Xmx30g) and the maximum stack size to 60M (-Xss60m). For other tools, we adopted the default tool configuration in our experiments. 
 \tool spent 262 seconds analyzing all programs.
As shown in Table~\ref{tab:32}, \tool outperformed the other tools by acquiring the highest average recall (\rec) and F-score (\fscore). It obtained the same average precision rate---\pre---as CryptoGuard and FindSecBugs, which rate is much higher than that of CogniCrypt (i.e., 58\%). 

\tool reported API misuses in eight projects, while the other three tools reported issues in nine projects. As the dataset labels no vulnerability in \codefont{tika.jar} and all tools found zero vulnerability in that project, we could not measure P, R, or F for these tools.
\tool was unable to analyze \codefont{tomee.jar}, because WALA does not always work well with the JAR files built by Maven~\cite{maven}. We believe that once WALA developers overcome the limitation between WALA and Maven JARs in the future, \tool can also analyze \codefont{tomee.jar}.
Among the four tools under comparison, CryptoGuard obtained a slightly lower F-score than \tool (78\% vs.~82\%), followed by FindSecBugs and CogniCrypt. Two possible reasons can explain \tool's higher F-score. First, thanks to its great extensibility, \tool has a larger pattern set of API misuses. Second, its inter-procedural analysis can accurately detect API misuses in more complex scenarios.

\emph{\textbf{Analysis of False Positives.}} 
We manually inspected the cases where \tool did not report misuses correctly. We found one reason to explain why \tool did not achieve 100\% precision: the ground truth is incomplete, as it labels some instead of all invocations of \codefont{Random()}. We currently consider the extra calls of \codefont{Random()} found by \tool to be false positives, although the actual precision rate is higher. 

\emph{\textbf{Analysis of False Negatives.}}
\tool missed some labeled API misuses, because the corresponding misuse patterns are not covered by our \templateCount inferred patterns. Some of these missing patterns can be added to \tool if we feed the tool with more code examples. One missing pattern cannot get added even if we provide $\langle I, S\rangle$ pairs to \tool. The pattern is to replace \codefont{new URL("http://...")} with \codefont{new URL("https://...")}. 
HTTPS is HTTP with encryption. Nowadays all websites are supposed to use HTTPS instead of HTTP for secure communication, so any URL string hardcoded in programs should always start with ``\codefont{https}'' instead of ``\codefont{http}''. In this pattern, the difference between insecure and secure code lies in the string literal, which is not handled by \tool currently. To derive the pattern from given code examples, we need to extend \tool and our specialized ways of example definition, to accurately locate and properly represent any difference within strings. 


\vspace{0.2em}
\noindent\begin{tabular}{|p{8.1cm}|}
	\hline
	\textbf{Finding 2:} \emph{On the third-party Apache dataset, \tool outperformed existing tools by achieving the highest precision, recall, and F-score on average. Our experiment indicates \tool's great capability of vulnerability detection. 
	}
\\
	\hline
\end{tabular}

 \begin{table}
\footnotesize
    \caption{The sampled vulnerabilities and repairs}\label{tab:manual}
    \vspace{-1.8em}
    \begin{tabular}{p{1.6cm}|R{.5cm}|R{.6cm}R{.5cm}R{.5cm}| R{1.2cm}R{.9cm}}
    \toprule
{\textbf{Security}}&{\textbf{\# of }}& \multicolumn{3}{c|}{\textbf{Vulnerability Detection}} & \multicolumn{2}{c}{\textbf{Repair Suggestion}} \\ \cline{3-7}
\textbf{Class API}& \textbf{Reports}& \textbf{Basic} & \textbf{Intra-}& \textbf{Inter-} & \textbf{Parameter or API replacement} 
&\textbf{Code replacement}    \\       
        \toprule
Cipher & 6 & 5 & & 1 &6 &\\
        \hline
          {HostnameVerifier} & 2 & 2 & & & &  2 \\ 
        \hline
          {IvParameterSpec} & 4 & & 1&3  & &4\\
        \hline
          {KeyPairGenerator} & 3  & & 1 & 2 &3\\
        \hline
          {KeyStore} & 5 & 1& &4 & &5\\
        \hline
          {MessageDigest} & 7 &5 & &2 &7 & \\
        \hline
          {PBEParameter- Spec} &7 &1 &4&2 &4&3 \\
        \hline
          {SecretKeyFactory} & 4&2&&2 &4\\
        \hline
          {SecretKeySpec} & 11& 6 & 1&4 & 9&8 \\
        \hline
          {SecureRandom}&5 &5& & &5& \\
        \hline
          {SSLContext} &5 & 5& & &5  \\
        \hline
          {TrustManager} &12&12&& &&12 \\
        \bottomrule
        \textbf{Total}  &71 &44&7&20 &43&34\\
        \bottomrule
    \end{tabular}
    \vspace{-2.5em}
\end{table}

\vspace{-.5em}
\subsection{Effectiveness of Repair Suggestion}
\label{sec:repair}

By applying \tool to the second dataset mentioned in Section~\ref{sec:evaluateApply}, we got hundreds of vulnerabilities reported together with repair suggestions.
Due to the time limit, we did not check every vulnerability as well as their repair(s); instead, we manually sampled the vulnerability reports and suggested repairs for \sampleCount misuse instances. 

To ensure the representativeness of our manual inspection results, we took four steps to create the sample set. First, we clustered all bug reports based on API-misuse patterns. 
Second, we randomly picked 10 reports from each cluster for further checking. If any cluster contained nine or fewer reports, we picked all reports. Third, when bug reports referred to duplicated code snippets, we removed duplicates to simplify our manual task, getting 71 sampled vulnerabilities. Fourth, we mapped the 71 samples to the security classes that they correspond to, and created Table 6. Notice that because some security classes (e.g., SecretKeySpec and TrustManager) have multiple method APIs that are prone to misuses (i.e., have multiple API-misuse patterns), the corresponding rows contain more than 10 samples (e.g., 11 for SecurityKeySpec and 12 for TrustManager). 

After the first author created the sample set, both the first and fifth authors independently checked bug reports and fixing suggestions. The two authors compared their manual inspection results for cross-checking; they had extensive discussion for any opinion divergence and even involved the second author into discussion, until reaching a consensus. 
  
Among the examined vulnerabilities, \tool revealed 44 misuses without doing any backward slicing (see the \textbf{Basic} column); it  successfully matched templates with single Java statements. \tool revealed seven misuses via intra-procedural slicing because in each of these scenarios, \tool located and analyzed multiple statements to match the related template. 
 \tool revealed 20 misuses via inter-procedural slicing, because multiple statements from different Java entities (i.e., methods or fields) demonstrate each of the misuses. 

70 of the vulnerabilities are true positives; the remaining one was falsely reported. This is because during its analysis, \tool checks whether the second parameter of \codefont{KeyStore.load(InputStream stream, char[] password)} is derived from a hardcoded constant; if so, the API call is considered insecure. Such analysis logic can effectively identify any password derived from a hardcoded secret. However, in our experiment, it incorrectly reported a scenario where the password is loaded from a file, whose name is hardcoded as a string literal. In the future, we will overcome this limitation by implementing heuristics (e.g., regular expressions) in \tool, to differentiate between constants serving for distinct purposes.
 
 
 Additionally, among the \repairCount suggested repairs, 43 repairs are solely about parameter/API replacement; \tool does not need to generate any code or customize any identifier to propose these fixes. Meanwhile, 34 repairs involve both multi-statement fixes and identifier customization. Notice that the total number of repair suggestions (i.e., \repairCount) is larger than the vulnerability count (i.e., 71). This is because \tool provides multiple suggestions for six vulnerabilities, as each of the code snippets matches two templates simultaneously and \tool suggests a repair for each match.  
Finally, we found all repairs by \tool to be correct. However, as \tool has a false positive when reporting API misuses, 
we count the repairs for the other 70 misuses as correct suggestions.

\vspace{0.2em}
\noindent\begin{tabular}{|p{8.1cm}|}
	\hline
	\textbf{Finding 3:} \emph{We manually checked \repairCount fixes generated by \tool, and found 76 of them to be correct. It indicates that \tool has great capability of repair suggestion.
	}
\\
	\hline
\end{tabular}

\vspace{-1.5em}
\section{Related Work}\label{sec:related} 

The related work of our research includes automatic detection of security-API misuses, and example-based program transformation.

\vspace{-.5em}
\subsection{Detection of Security-API Misuses}
Tools were built to detect security-API misuses~\cite{fahl2012eve,egele2013empirical,he2015vetting,ma2016cdrep,ma2017vurle,kruger2017cognicrypt,kruger2018crysl,Fischer2017,Rahaman2019,findsecbugs,sonarqube}.
As shown in Table~\ref{tab:compare}, most tools 
statically analyze programs based on hardcoded or built-in rules.
Specifically, CryptoLint hardcoded six API misuse patterns. For each located potentially vulnerable API call (e.g., \codefont{Cipher.getInstance(v)}), CryptoLint conducts backward slicing to decide whether the used parameter value is insecure (e.g., \codefont{v="AES/ECB"}). CDRep reimplements the design of CryptoLint for misuse detection. It also repairs detected misuses leveraging  manually created patch templates. Such tools are not easy to extend, because tool builders or users have to modify tool implementation to expand the rulesets of vulnerabilities.

Fischer et al.~\cite{Fischer2017} built a tool to detect misuses in two ways: machine learning and graph matching. Both methods detect vulnerabilities based on the similarity between given programs and labeled (in)secure code. 
 However, this tool does not rigorously reason about misuse patterns; it cannot pinpoint the exact location of misused API in vulnerable code.
CogniCrypt~\cite{kruger2017cognicrypt} supports rule definition via a domain-specific language CrySL~\cite{kruger2018crysl}. Each CrySL rule 
specifies \textbf{correct} API usage, and 
CogniCrypt detects misuses by scanning programs for rule violation. 
CogniCrypt has three limitations. First, manually prescribing rules with CrySL can be tedious and error-prone for tool users. Second, CrySL cannot express the API misuses related to constant placeholders and  constants within certain ranges. 
Third, CogniCrypt does not customize fixes.

VuRLE~\cite{ma2017vurle} is most relevant to our work. VuRLE also detects and repairs vulnerabilities based on the $\langle I, S\rangle$ code examples provided by users. 
 \tool complements VuRLE in three ways. First, \tool infers each pattern from one instead of multiple code pairs, so it works well when users have only one pair.
Second, \tool conducts inter-procedural analysis and adopts succinct info (security APIs and data dependencies) to match code with templates, while VuRLE uses intra-procedural analysis and tree matching. Thus, \tool can find more matches. Listing~\ref{lst:p} is an exemplar program, where the API misuse is only identifiable when a tool conducts inter-procedural program analysis. VuRLE cannot locate the API misuse but Seader can. 
   Third, \tool supports specialized example definitions, and VuRLE does not.    
   As the source code of VuRLE is not publicly available, we cannot empirically compare it with \tool.   

\vspace{-0.5em}
\subsection{Example-Based Program Transformation}

Based on the insight that developers modify similar code in similar ways, researchers built tools to infer program transformations from exemplar code change examples, and to manipulate code or suggest changes accordingly~\cite{Miller2001,Pham2010,Meng2011:sydit,Meng2013:lase,rolim2017learning,An2018,Xu2019}. For instance, 
given one or multiple code change examples, LASE~\cite{Meng2013:lase} and REFAZER~\cite{rolim2017learning} infer a program transformation from the examples; they then use the transformation to locate similar code to edit, and apply customized transformations to those locations.
Given a set of vulnerable and patched code fragments $K=\{(A_1, A'_1), (A_2, A'_2), \ldots, (A_n,$ $A'_n)\}$, SecureSync~\cite{Pham2010} scans the source code of programs to find fragments, which are similar to vulnerable code $A_i$ but dissimilar to the patched code $A_i' (i\in[1, n])$. 

Sharing the same insight, we designed \tool to detect and fix vulnerabilities based on code examples. However, \tool is different from prior work for two reasons. First,  \tool infers a program transformation via \emph{intra-procedural} analysis, but conducts pattern matching via \emph{inter-procedural} analysis. All the tools mentioned above are limited to intra-procedural analysis. Our unique design makes \tool more powerful when it searches for pattern matches; it can find matches that go beyond the method boundary and span multiple Java methods. Second, \tool supports three specialized ways of example specification, which can describe transformations that are not expressible via plain code examples. Based on our experience with security-API misuses, these unique specification methods are necessary and helpful.

\vspace{-0.5em}
\section{Threats to validity}
\label{sec:threats}
 All inferred patterns and detected vulnerabilities are limited to our experiment datasets and two cryptographic libraries: JCA and JSSE. The observations may not generalize well to other subject programs (e.g., closed-source projects) or other security libraries.
 We actually also manually checked API misuses in Spring Security---a widely-used third-party security framework, and found more misuse patterns that can be handled by \tool (e.g., the parameter value of BCryptPasswordEncoder's constructor should not be less than 10).
 \tool can handle the API misuses that involve (1) calling certain method APIs with incorrect parameter values, (2) calling certain method APIs in incorrect sequential orders, and (3) incorrectly overriding certain method APIs. Therefore, \tool is generalizable in terms of (1) the API misuse patterns to handle, and (2) security libraries to cover.



In some repair suggestions provided by \tool, there are placeholders that we need developers to further customize (see ``\codefont{//Please change `example.com' as needed}'' in Figure~\ref{fig:example3}). Such placeholders should be filled based on developers' software environments, or even require extra configurations outside the codebase (e.g., generating and loading SSL certificates). 
In the future, we will provide clearer suggestions on hands-on experience and create interactive tools that guide developers to apply complete repairs step-by-step. 

\vspace{-0.5em}
\section{Conclusion}
We created \tool---a new approach to take in  $\langle$insecure, secure$\rangle$ code examples, infer vulnerability-repair patterns from examples, and apply those patterns for vulnerability detection and repair suggestion. Compared with prior work, \tool offers a more powerful means for security experts to extend the pattern set of API-misuse detectors,  
 and concretizes security expertise as customized fixing edits for developers.
Our evaluation shows \tool's great capabilities of pattern inference and application; it detects API misuses and suggests fixes with high accuracy. 
In the future, we will widen \tool's applicability by specifying more code pairs. We will also extend \tool's capability by adding support for more kinds of API misuse patterns (e.g., patterns involving Java annotations).

\vspace{-.5em}
\section*{Acknowledgments}
We thank anonymous reviewers and Dr. Eric Bodden for their valuable comments. This work was supported by NSF-1845446 and NSF-1929701.
\bibliographystyle{ACM-Reference-Format}
\bibliography{reference}

\end{document}